# New method for determining avalanche breakdown voltage of silicon photomultipliers

Igor Chirikov-Zorin[*]

Joint Institute for Nuclear Research, Dubna, Russia

The avalanche breakdown and Geiger mode of the silicon *p-n* junction is considered. A precise physically motivated method is proposed for determining the avalanche breakdown voltage of silicon photomultipliers (SiPM). The method is based on measuring the dependence of the relative photon detection efficiency ($PDE_{rel}$) on the bias voltage when one type of carriers (electron or hole) is injected into the avalanche multiplication zone of the *p-n* junction. The injection of electrons or holes from the base region of the SiPM semiconductor structure is performed using short-wave or long-wave light. At a low overvoltage (1–2 V) the detection efficiency is linearly dependent on the bias voltage; therefore, extrapolation to zero $PDE_{rel}$ value determines the SiPM avalanche breakdown voltage with an accuracy within a few millivolts.

## 1. Introduction

In modern experimental physics, development of new devices shows a clear trend to replace traditional optical detectors by silicon photomultipliers (SiPMs) – a new generation of photodetectors. SiPMs are micropixel silicon avalanche photodiodes (APD) operating in the so-called Geiger mode at a bias voltage $V_b$ higher than the breakdown voltage $V_{br}$. Therefore, SiPMs have a high internal amplification rate $\sim 10^6$ at room temperature and can detect weak light fluxes at a level of individual photons counts.

---

[*] E-mail address: chirikov@jinr.ru

The SiPM is a rectangular matrix of identical small square APD pixels with quenching resistors connected in parallel on the surface of a common silicon substrate. The typical pixel density is $10^2 - 10^4$ mm$^{-2}$ with a pitch of $10-100$ μm.

The pixels are independent photon microcounters working in the digital (Geiger) "yes/no" mode. They generate a standard signal when detecting a single photon, but at the detection flashes of light, when many pixels are simultaneously fired, the output signal on the total load is a sum of standard signals. Thus, the SiPM is in general an analog device capable of measuring the intensity of light with a dynamic range corresponding to the total number of pixels.

The main characteristics of the SiPM (gain, detection efficiency, and others) are determined by the overvoltage $\Delta V = V_b - V_{br}$. Therefore, the avalanche breakdown voltage $V_{br}$ is an important parameter, the determination of which is necessary for detailed studies and application of the SiPMs. The spread of $V_{br}$ allows estimating the influence of different technological factors and processes in the development and production of the SiPMs.

Note that $V_{br}$ increases with temperature. The temperature coefficient of the avalanche breakdown voltage $V_{br}(T)$ is also an important parameter of SiPMs.

There are several methods for determining the breakdown voltage of the SiPMs [1-11], which are based on different assumptions and models and therefore have limited accuracy. In this paper present the new precision method for determining $V_{br}$.

## 2. The gain, breakdown voltage and Geiger mode of an APD

The internal gain in the photodiode can be obtained, in principle, by creating a strong electric field in the depleted region of the *p-n* junction and free charge carriers accelerated in this field could produce impact ionization of the *p-n* junction and the develop the avalanche multiplication process. The dependence of the gain $G$ on the reverse bias voltage $V_b$ of the avalanche photodiode and different regions of the electric discharge by analogy with the operating modes of the gas counter of nuclear radiation are schematically shown in Fig. 1.

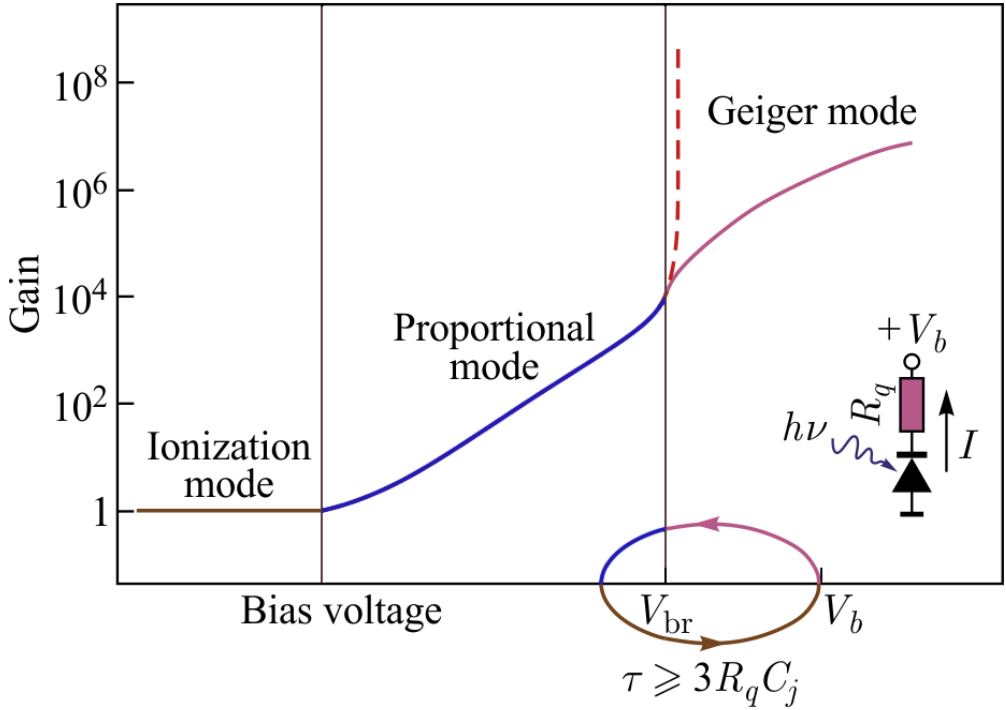

Fig. 1. The schematic representation of the gain versus the reverse bias voltage for an avalanche photodiode.

The multiplication process of free carriers in the *p-n* junction is characterized by a *k*-factor equal to the ratio of the impact ionization coefficients of holes and electrons ($k=\beta/\alpha$). For silicon, $\alpha>\beta$ for any values of the electric field strength [12,13]. In the proportional region (i.e., when the output signal is proportional to the intensity of the detected light) avalanche multiplication is mainly due to electrons ($k \ll 1$) and therefore the avalanche is quenched. The gain of the avalanche photodiode in the proportional mode

$$G_e \approx \exp(\alpha W), \quad (1)$$

where $W$ is the width of the avalanche multiplication zone and reaches $10^4$ at room temperature [14].

With increasing $V_b$ the electric field strength in the depletion region of the *p-n* junction grows, which leads to an increase in the impact ionization coefficients of electrons and holes, and k-factor [13]. The avalanche multiplication at a bias voltage higher than a certain threshold value, so-called breakdown $V_{br}$, is due to both electrons and holes ($k\rightarrow 1$). At high electric fields ($E > 10^5$ V/cm) impact ionization of the *p-n* junction by holes gives rise to positive internal feedback and, as a consequence, to development of self-sustaining electron-hole avalanche and thermal breakdown of the photodiode *p-n* junction [15]. At the bias voltage $>V_{br}$ production of one electron-hole pair in the avalanche multiplication region of the *p-n* junction is enough to initiate a self-sustaining discharge with a formal gain $G_{eh}\rightarrow\infty$. The probability of triggering the avalanche breakdown by an electron or a hole is given by $P_{pair}=P_e+P_h-P_e P_h$, where $P_e$ and $P_h$ are the electrons and holes triggering probabilities. It is important to note that when $V_b=V_{br}$, the triggering probability is $P_e=P_h=0$.

The discharge can be interrupted by a quenching resistor $R_q$ connected in the bias circuit in series with a photodiode (Fig. 1), which creates a negative external feedback. To quench the discharge, it is necessary to reduce on the *p-n* junction to voltage $V_{br}$, i.e., the voltage drop across the resistor during the discharge should be equal to or larger than the so-called overvoltage $\Delta V=V_b-V_{br}$. Thus, for quenching the discharge the condition $R_q I \geq V_b-V_{br}$ should be fulfilled. The electron-hole avalanche stops when $V_b=V_{br}$, but the voltage across the photodiode continues to decrease as a result of draining of the accumulated charge carriers from the *p-n* junction to the base regions of the APD.

The time of recovery of the *p-n* junction potential to its original condition after the avalanche discharge is $\tau \geq 3R_q C_j$ in accordance with the exponential law of the recharging of the *p-n* junction capacitance $C_j$ through the quenching resistor $R_q$. With increasing bias voltage, the recovery time significantly increases as a result of partial discharge of $C_j$ due to the increase in the dark current, and others [16].

Thus, the APD operation at a bias voltage higher than the breakdown voltage with negative feedback made via the resistor $R_q$ in the bias circuit in series with the photodiode is called the Geiger mode by analogy with the gas-discharge Geiger-Muller counter. Due to the linear element $R_q$ in connected in the bias circuit, the gain in the Geiger mode is a linear function of the overvoltage.

In conclusion, it should be noted that during transition from the proportional[*] to the Geiger mode ($\Delta V = V_b - V_{br} \to 0$), the SiPM gain abruptly increases and reaches $\sim 10^4$, but is not equal to zero, as it is assumed in the method for determining $V_{br}$ proposed in Refs. [1,4,5,7,8,10] and others.

## 3. The method for determining avalanche breakdown voltage of SiPMs

The new physically motivated precision method for determining $V_{br}$ is based on measurements of the relative SiPM detection efficiency.

The photon detection efficiency (*PDE*) is the product of the geometrical factor $\varepsilon$ (ratio of sensitive area to total surface of SiPM), pixel quantum efficiency (*QE*), and probability $P_{tr}$ for triggering avalanche breakdown by the produced free charge carriers

$$PDE = \varepsilon \, QE(\lambda) \, P_{tr}(\lambda, T, \Delta V). \qquad (2)$$

The quantum efficiency of the pixel is determined by the probability for production of the charge carriers in the sensitive volume and significantly depends on the wavelength $\lambda$ of the detected light

$$QE(\lambda) = [1-r][\exp(-L_{ep}/\lambda_{ep}(\lambda))][\exp(-L_{SiO_2}/\lambda_{SiO_2}(\lambda))][1-\exp(-L_{Si}/\lambda_{Si}(\lambda))], \qquad (3)$$

where r is the reflection coefficient from the pixel surface, $L_{ep}$ and $L_{SiO_2}$ are the thicknesses of the epoxy resin and silicon oxide layers used as protective coatings, $\lambda_{ep}(\lambda)$ and $\lambda_{SiO_2}(\lambda)$ are the light absorption lengths in the coatings, $L_{Si} = W_n + W + W_p$ is the thickness of the sensitive layer (Fig. 2), and $\lambda_{Si}(\lambda)$ is the light absorption length in silicon.

---

[*] The gain of the SiPM in the proportional mode ($V_b < V_{br}$) is determined by expression (1), and is small due to the narrow width of the avalanche multiplication zone (Fig. 2).

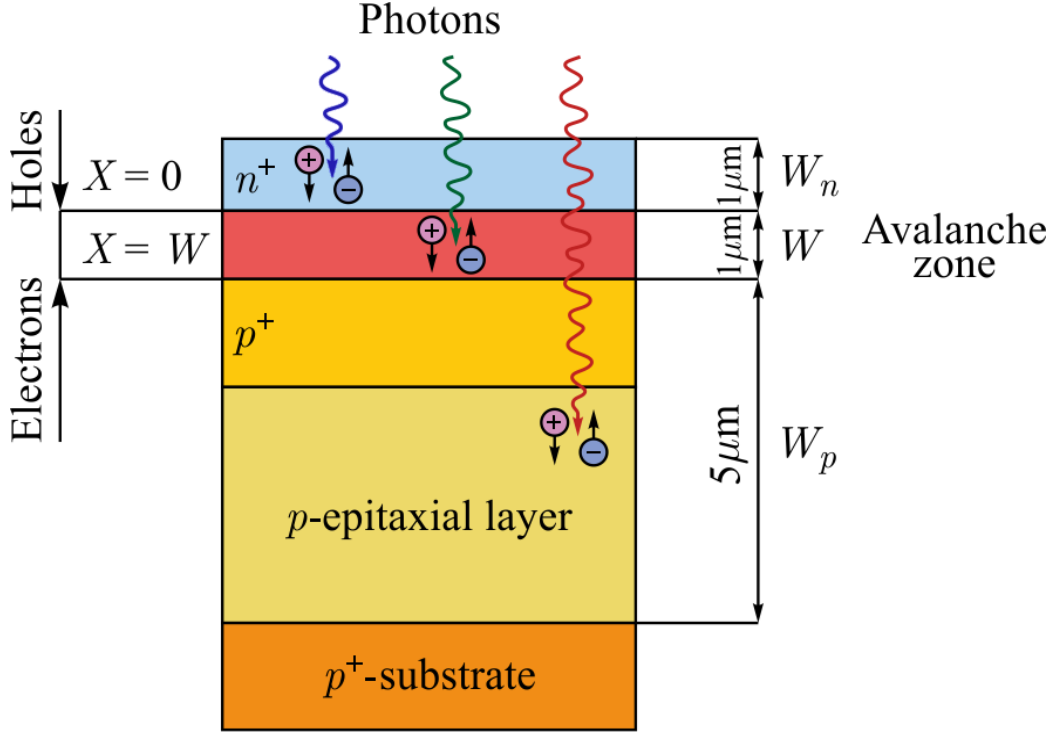

Fig. 2. A simplified schematic view of the pixel structure on the *p*-substrate.

The avalanche breakdown triggering probability depends on the overvoltage and on the position where the primary electron-hole pair is produced in the sensitive layer of the pixel. Three photogeneration zones of primary electron-hole pairs $W_n$, $W$, $W_p$ and their typical thickness [17] are shown in Fig. 2. The triggering probability of the electron-hole pair in the avalanche multiplication zone $W$ is given by

$$P_{\text{pair}}(x, \Delta V) = P_e(x, \Delta V) + P_h(x, \Delta V) - P_e(x, \Delta V) P_h(x, \Delta V), \qquad (4)$$

where $P_e(x, \Delta V)$ and $P_h(x, \Delta V)$ are the triggering probability for electrons and for holes generated at the point with the coordinate $x$ at overvoltage $\Delta V$. Figure 3 shows the dependence of $P_e(x, \Delta V)$ and $P_h(x, \Delta V)$ on the coordinates of the generation of charge carriers in the avalanche multiplication zone calculated by W.G. Oldham et al. [18].

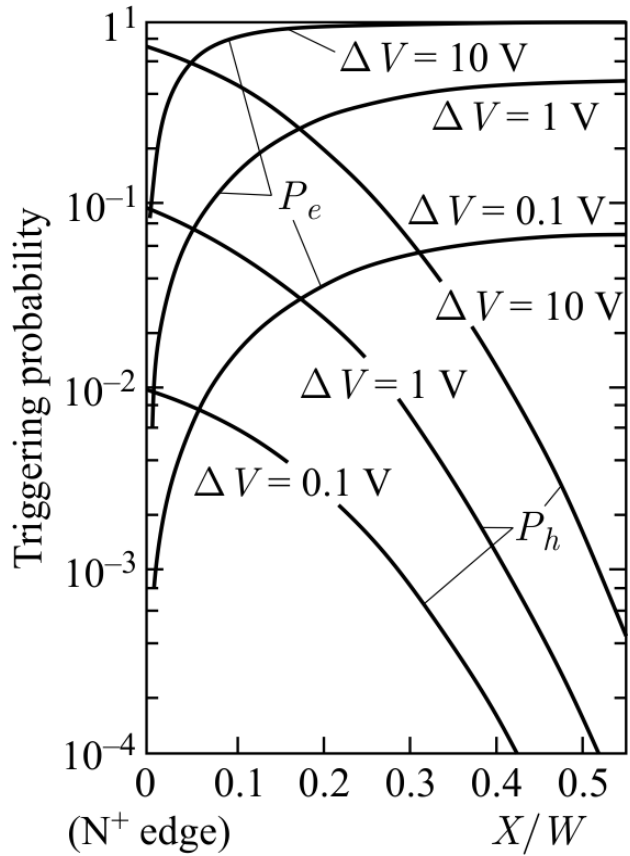

Fig. 3. The triggering probabilities $P_e(x, \Delta V)$ and $P_h(x, \Delta V)$ as a function of the positions of the carriers generated in the avalanche multiplication zone with a width $W$ at several overvoltage values $\Delta V$. Coordinates of the zone boundaries $X = 0$ and $X = W$ are shown in Fig. 2.

Electrons and holes produced in the zone $W_p$ drift in the electric field in opposite directions, holes to the *p*-base and electrons to the avalanche multiplication zone $W$, an thus the latter can triggering an avalanche breakdown. On the contrary, if the photons are absorbed in zone $W_n$, an avalanche can be triggered only by holes. The avalanche triggering probability of electrons and holes produced outside multiplication zone $W$ was calculated and verified experimentally by P. Antognetti and W.G. Oldham (Fig. 4) [19]. The triggering probability of holes is much lower than that of electrons due to the difference between of their impact ionization coefficients ($\alpha > \beta$).

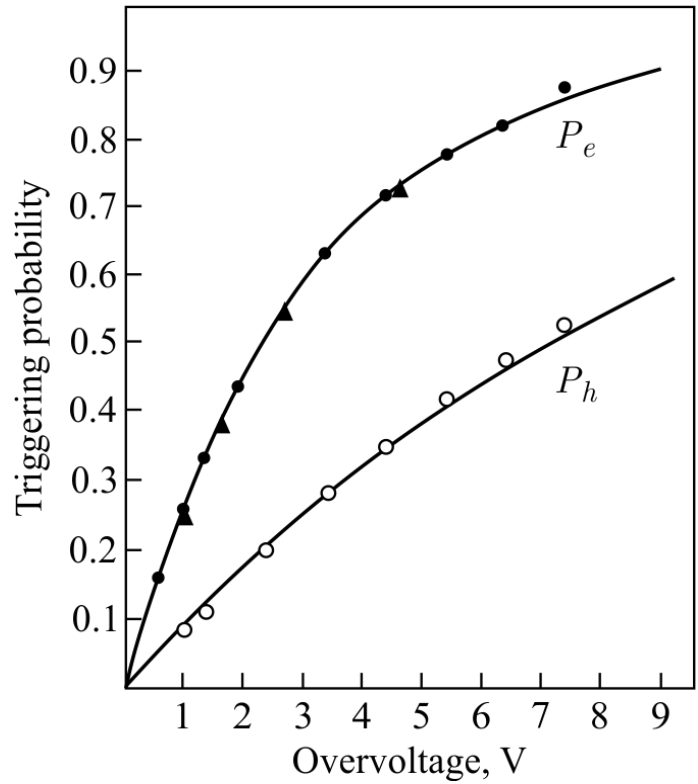

Fig. 4. The avalanche triggering probabilities for electrons $P_e$ and for holes $P_h$ as a function of the overvoltage $\Delta V$.

As can be seen from Fig. 4, the triggering probabilities of electrons and holes injected into the avalanche multiplication zone of the *p-n* junction are linear functions of overvoltage at its low values (1–2 V). The *PDE* is proportional to the triggering probability; therefore, it is also a linear function of the overvoltage and consequently of the bias voltage $V_c$. Extrapolation of the function $PDE = PDE(V_c)$ or relative $PDE_{rel} = PDE_{rel}(V_c)$ to zero values determines the SiPM avalanche breakdown voltage.

Electrons or holes can be injected into the *W* avalanche multiplication zone using short-wave or long-wave light. If the SiPM is illuminated by short-wave light with $\lambda = 400$ nm, it is almost completely absorbed (>99%) in the layer $W_n$, and holes will drift into the zone *W*. If illumination is made by long-wavelength light with $\lambda = 1000$ nm, which is weakly absorbed in thin of $W_n$ and *W* layers ($\approx 1\%$), it is mainly electrons that will drift from the layer $W_p$ into the avalanche multiplication zone.

## 4. The measurement and results

To illustrate the method, we determine the avalanche breakdown voltage for the SiPMs of two types SSPM-050701GR-TO18 and S60 [20]. Their *PDE* is measured by the method of low-intensity light flashes (~10 photons) method [21,22], when the probability for production of free charge carriers in the sensitive layer of SiPM is low.

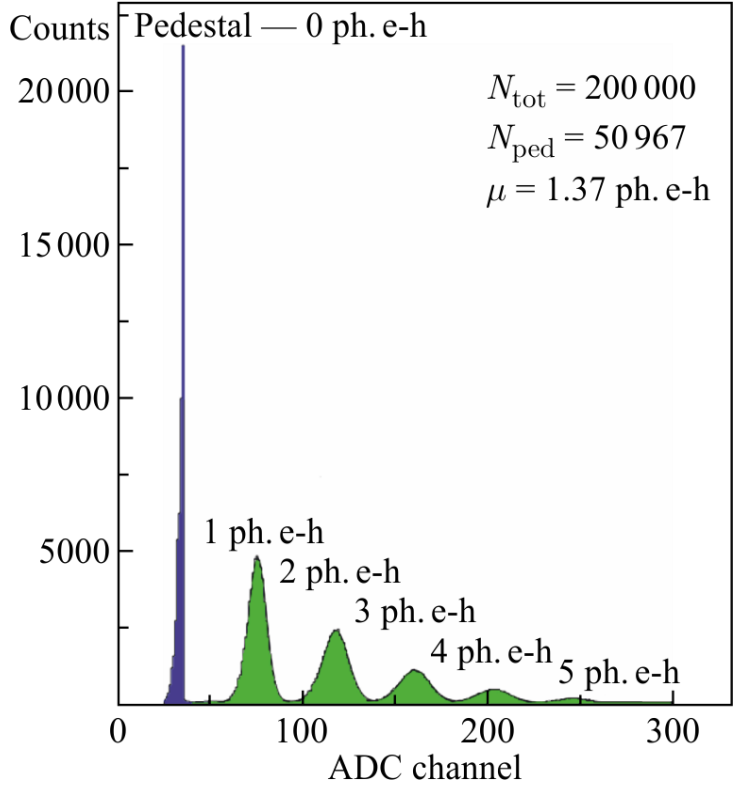

Fig. 5. The S60 SiPM pulse height spectrum for low-intensity light flashes with the mean number of photogenerated electron-hole pairs µ = 1.37 ph. e-h.

A typical LED charge spectrum of the SiPM for low-intensity light pulses is shown in Fig. 5. In accordance with the Poisson distribution $P(n;\mu) = \mu^n \exp(-\mu)/n!$, the average number of photogenerated pairs, that triggered avalanche breakdown pixels of the SiPM is

$$\mu = -\ln P(0;\mu), \qquad (5)$$

where $P(0;\mu) = N_{ped}/N_{tot}$ is the probability of non-generation of a signal from SiPM, $N_{ped}$ is the number of events in the pedestal, and $N_{tot}$ is the total number of events in the spectrum. The photon detection efficiency was determined from the expression

$$PDE = -\ln(N_{ped}/N_{tot})/N_{ph}, \qquad (6)$$

where $N_{ph}$ is the average number of photons incident on the SiPM. The average number of photons was estimated by the H6780-04 photosensor (Hamamatsu) with the known quantum efficiency using the low-intensity light flash method.

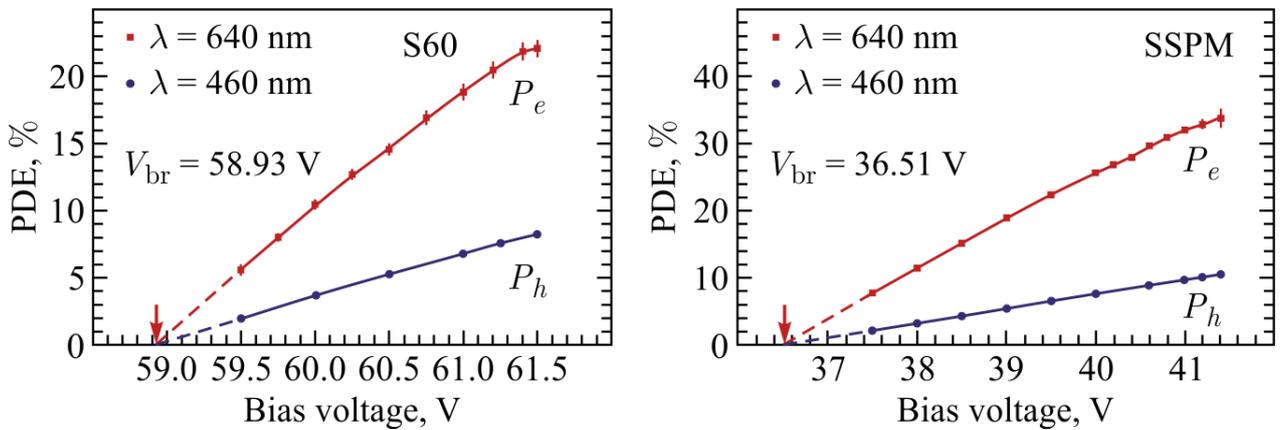

Fig. 6. The photon detection efficiency as a function of the overvoltage for different wavelengths of light for the S60 and SSPM-050701GR-TO18 SiPMs based on $p$-substrate.

The *PDE* measurement were performed using red ($\lambda \approx 640$ nm) and blue light ($\lambda \approx 460$ nm), when mainly one type of free charge carriers, electrons or holes are injected in the avalanche multiplication region. The linear extrapolation of the initial part of $PDE = PDE(V_c)$ dependence to zero values determines the SiPM avalanche breakdown voltage (Fig. 6).

## 5. Conclusion

The values of breakdown voltage obtained for the SiPMs of two types S60 and SSPM-050701GR-TO18 for different wavelengths of light coincide with high accuracy $\sim 10^{-3}$ V, which confirms the reliability of the proposed method.